# Reworked Second Order Blind Identification and Support Vector Machine technique towards imagery movement identification from EEG signals

Kalogiannis Gregory, *Member, IEEE*, Hassapis George, *Senior Member, IEEE*

*Abstract—* During imagery motor movements tasks, the so called mu and beta event related desynchronization (ERD) and synchronization (ERS) are taking place, allowing us to determine human patient imagery movement. However, initial recordings of electroencephalography (EEG) signals contain system and environmental noise as well as interference that must be ejected in order to separate the ERS/ERD events from the rest of the signal.

This paper presents a new technique based on a reworked Second Order Blind Identification (SOBI) algorithm for noise removal while imagery movement classification is implemented using Support Vector Machine (SVM) technique.

## I. Introduction

Motor rehabilitation devices, such as continuous passive motion machines (CPM), are used for rehabilitation in clinics, hospitals or physiotherapists and they are vital to therapeutic and restorative treatment [1]. These devices allow their easily connection with controllers in order to follow trajectories determined by the patient's intensions and will for movement.

Brain signals produced by human intention for movement can be used for controlling motors in such devices. In our previous work [2], an EEG process technique, based on identification ERD/ERS [3] phenomena during motor imagery, using the Support Vector Machine (SVM) technique, was proposed, as showed in Figure 1.

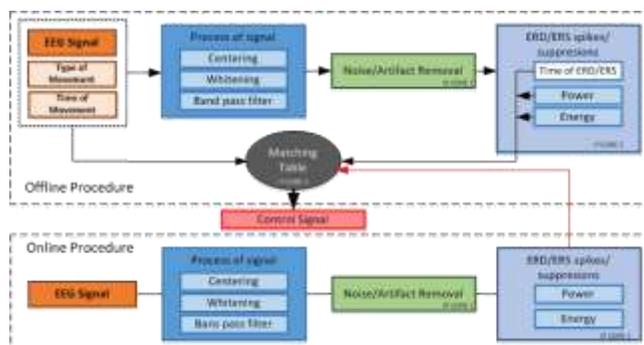

Figure 1. The process technique of EEG based on ERD/ERS events

Implementation of such an architecture requires fast recognition of the motor imagery movements of the joint in order to create the appropriate control signals. This can be done by manipulating the EEG data with the purpose of removing the noise and information that is not essential for creating the control signal in a fast and effective way. A new, SOBI algorithm based on SCHUR decomposition for removing artifacts is proposed in this study.

## II. Noise Removal

SOBI independent component analysis (ICA) algorithm is a widely used algorithm in removing artifacts from EEG signals. Using different diagonalization method such as the SCHUR decomposition we can reduce the time of noise removal from EEG sources. The evaluation of the EEG processing based on SOBI and modified SOBI algorithm are shown in Table 1.

TABLE I. Execution Time Of Reworked And Typical SOBI Blind Source Seperation Algorithm, In Seconds

| Dataset | Execution time of EEG processing algorithm in seconds | |
|---|---|---|
| | *based on modified SOBI* | *based on typical SOBI* |
| 1 | 1.567458 | 9.845235 |
| 2 | 1.845277 | 10.379097 |
| 3 | 1.814460 | 11.514513 |
| 4 | 1.739605 | 9.800095 |
| 5 | 1.600469 | 9.799495 |

## III. Signal Classification

In order to determine the motor imaginary movement, features such as power and energy of the signal obtained after the removal of artifacts are calculated. These features are compared with the respectively same features of classified signals from an available database while trying to identify the class to which the processed signal belongs. The database classification is done off-line by using the SVM algorithm.

Kalogiannis Gregory is with the Electrical Engineering Department, Aristotle University of Thessaloniki, Greece (e-mail: gkalogiannis@ece.auth.gr).

Hassapis George is with the Electrical Engineering Department, Aristotle University of Thessaloniki, Greece (e-mail: chassapis@eng.auth.gr).